# Effects of Interfacial Oxygen Diffusion on the Magnetic Properties and Thermal Stability of Pd/CoFeB/Pd/Ta Heterostructure

Saravanan Lakshmanan[1, *], Cristian Romanque[1], Mario Mery[1], Manivel Raja Muthuvel[2], Nanhe Kumar Gupta[3], and Carlos Garcia[1, *]

[1]*Departamento de Física, Universidad Técnica Federico Santa María, 2390123 Valparaíso, Chile*
[2]*Advanced Magnetics Group, Defence Metallurgical Research Laboratory, Hyderabad-500 058, India*
[3]*Thin Film Laboratory, Department of Physics, Indian Institute of Technology Delhi, New Delhi 110016, India*

[*]E-mail: saravanan.lakshmanan@usm.cl; carlos.garcia@usm.cl

## Abstract

We investigated the effects of annealing temperatures ($T_A$) on a Pd (5 nm)/CoFeB (10 nm)/Pd (3 nm)/Ta (10 nm) multilayer structure. The as-deposited sample showed an amorphous state with in-plane uniaxial magnetic anisotropy (UMA), resulting in low coercivity and moderate damping constant ($\alpha$) values. Increasing $T_A$ led to crystallization, forming bcc-CoFe (110) crystals, which increased in-plane coercivity and introduced isotropic magnetic anisotropy, slightly reducing the $\alpha$. The two-fold UMA persists up to 600ºC, and the thermal stability of the in-plane magnetic anisotropy remains intact even $T_A$ = 700ºC. The $T_A$ significantly influenced the magnetic properties such as in-plane saturation magnetization ($M_{s//}$), in-plane and out-of-plane coercivities ($H_{c//}$ and $H_{c\perp}$), and in-plane effective magnetic anisotropy energy density ($K_{eff}$). Above 600ºC, $K_{eff}$ decreased, indicating a transition towards uniaxial perpendicular magnetic anisotropy. Interfacial oxidation and diffusion from the Ta capping layer to the Pd/CoFeB/Pd interfaces were observed, influencing chemical bonding states. Annealing at 700ºC, reduced oxygen within $TaO_x$ through a redox reaction involving Ta crystallization, forming TaB, PdO, and $BO_x$ states. Ferromagnetic resonance spectra analysis indicated variations in resonance field ($H_r$) due to local chemical environments. The $\alpha$ reduction, reaching a minimum at 300ºC annealing, was attributed to reduced structural disorder from inhomogeneities. Tailoring magnetic anisotropy and spin dynamic properties in Pd/CoFeB/Pd/Ta structures through $T_A$-controlled oxygen diffusion/oxidation highlights their potential for SOT, DMI, and magnetic skyrmion-based spintronic devices.

# 1. INTRODUCTION

To achieve a giant tunneling magnetoresistance (TMR) ratio, numerous efforts have been dedicated to improving spintronic devices, leading to the identification of many structures with new phenomena [1,2]. One of the most crucial device configurations is magnetic tunnel junctions (MTJs), consisting of reference and free layers separated by a few nanometers of an insulating layer. The spin orientation of the reference layer is generally fixed, while the spin orientation of the free layer of the ferromagnet is considered a storage layer, and its magnetization is freely switched by methods such as an external magnetic field, spin-transfer torque (STT), and spin-orbit torque (SOT) [3–5]. Among these, SOT-based devices have two major advantages compared to STT-based devices, especially in magnetic random-access memory (MRAM). Firstly, SOT switching (≈ 1 ns) is significantly faster than STT switching (≈ 10 ns). Secondly, since reading and writing paths are different in SOT-MRAM, more stable devices are expected [6]. Heavy metal (HM) and ferromagnetic (FM) bilayer structures significantly fulfill the requirements of SOT-based devices. Moreover, FM/HM heterostructures strongly promote phenomena of spin-orbit interaction (SOI), including magnetic anisotropy in both in-plane and out-of-plane directions [7–9], antisymmetric exchange between neighboring 3d metals [10–12], topological spin textures such as bimerons, skyrmioniums, and skyrmions [13,14], chiral damping of magnetic domain walls [15], SOT [16], Spin Hall effect [17], and additional spin-related effects [18, 19]. More precisely, these phenomena are applicable in real-life applications, including the demanding improvement of skyrmions racetrack memories [20], neuromorphic computing [21, 22] and logic devices [23], SOT-based nano oscillators [24], and field-free induced magnetization switching [25, 26].

Magnetic anisotropy is a fundamental property of magnetic materials and serves as the cornerstone of spintronic devices. Anisotropy establishes an energy barrier between two opposing spin orientations, enabling the processing and storage of information. [27]. Different types of magnetic alloy films, including Co(Ta,Nb)Zr, FeAlSi, FeTaN, FeTaC, FeCoNi, CoNbFe, and CoFeB, both in single and multilayer structures, have been studied to enhance their magnetic properties. The emphasis lies on achieving tunable magnetic anisotropy, high saturation magnetization, moderate coercivity, and other desirable characteristics [28-33]. Among them, CoFeB-based alloys stand out as potential candidates for spintronic devices due to their large TMR of 1056% at room temperature [34]. Numerous reports have been published on CoFeB to understand the structure of domains [35], perpendicular magnetic anisotropy (PMA) [36], low Gilbert damping constant ($\alpha$) [37], and high spin polarization (P) [38]. Notably, thick CoFeB films with ultra-low $\alpha$ are particularly beneficial for low-power-consuming spintronic devices [28, 37]. The presence of in-plane or out-of-plane UMA in CoFeB films can be influenced by the type of multilayer structure, thermal treatment, and film thickness [36, 39, 40]. For modern spintronic devices, HM/FM and/or FM/HM-based multilayer structures are strongly required for both in-plane and out-of-plane magnetic tunnel junctions (i and p-MTJs) to evaluate magnetization switching. The magnitude of STT and SOT switching current can be diminished by decreasing $\alpha$ and increasing the spin Hall angle ($\theta_{SHA}$), respectively. Moreover, HM and FM-based structures regulate the DMI effect, influencing the manipulation of magnetic skyrmions [41]. The nonmagnetic buffer and over-layers with a desired face-centered cubic (fcc) structure are essential for tuning the strong magnetic anisotropy in CoFeB/Pd structures [40, 42-45]. In particular, the Pd layer has received great attention in spin-orbit torque (SOT) switching research due to its cost-effectiveness, ease of achieving PMA, and its utility as a reliable gas sensing material, among other factors [46, 47]. Moreover, Pd and PdO are thermodynamically stable [48], achieving magnetic anisotropy in CoFeB/Pd bilayer films to withstand higher annealing temperatures. For developing magnetic tunnel junction (MTJ) structures, the thermal stability of multilayer films should be satisfactory for industrial standard processing temperatures of 300ºC and 400ºC [49, 50]. On the other hand, Tantalum (Ta) plays a crucial role, easily helping to tune in-plane or out-of-plane magnetic anisotropy, offering high $\theta_{SHA}$, having good affinity with Boron atoms, and being widely used as a buffer and/or capping layer in MTJ-based PMA structures. Additionally, the spin-pumping effect in the Ta film cannot be ignored, as it acts as a good spin scatterer for a spin current created by spin precession in the CoFeB film [5, 28, 51-56]. Therefore, in the present study, Ta was used as a capping layer, making it easily suitable for manufacturing SOT-based MRAMs.

Here, we report a new type of multilayer structure with Pd/CoFeB/Pd/Ta with a thick magnetic layer, featuring UMA with two-fold symmetry, low magnetization damping and higher thermal stability of the UMA. Structural and magnetic properties, including saturation magnetization ($M_s$), coercivity ($H_c$), anisotropy energy density ($K_{eff}$), and damping parameters were thoroughly examined to elucidate the impact of thermal annealing ($T_A$) on UMA, $K_{eff}$, boron and oxygen diffusion, and the magnetization dynamics. This comprehensive analysis provides crucial insights for the development of novel spintronic devices, offering perspectives from both fundamental and practical applications.

## 2. EXPERIMENTAL SECTION

### *2.1 Preparation of the Multilayers*

Pd(5 nm)/Co$_{40}$Fe$_{40}$B$_{20}$(10 nm)/Pd(3 nm)/Ta(10 nm) multilayer films were fabricated on thermally oxidized Si (Si/SiO$_2$) substrates at room temperature (RT) using an Ultra-High Vacuum (UHV) Direct Current (DC)/Radio Frequency (RF) magnetron sputtering system from *AJA International, Inc., USA*. The deposition process was conducted under a 5.8×10$^{-9}$ Torr base pressure. Co$_{40}$Fe$_{40}$B$_{20}$(CoFeB) and Pd were DC sputtered at 40 W and 3 mTorr Ar, resulting in growth rates of 0.16 Å/s and 0.68 Å/s, respectively. Ta was RF sputtered at 80 W in 3mTorr Ar, with a growth rate of 0.34 Å/s. Growth rates were monitored using a Quartz Crystal Monitor (QCM) integrated into the UHV sputtering system. The thickness of each layer was confirmed using X-ray reflectivity (XRR) with a *PANalytical X¨pert PRO* and Atomic Force Microscopy (AFM) from *Nanomagnetic Instruments*. During layer deposition, all the substrates were rotated at up to 50 RPM to ensure the formation of a homogeneous film. To protect the CoFeB-based structures from external impurities and oxidation, thick Ta layers were used as a capping layer. After deposition, the *ex-situ* annealing process was conducted in the UHV main chamber under 1.5 × 10$^{-6}$ Torr at various annealing temperatures, reaching up to 700°C for 30 minutes without the presence of a magnetic field. The temperature was ramped at a rate of 2°C/s, and the samples were allowed to cool down slowly to RT before removal.

### *2.2 Characterizations*

The crystallographic orientation of the stacks was probed using a grazing incidence X-ray Diffractometer (GI-XRD) *[PANalytical X¨pert PRO]* equipped with a Cu anode [Cu K$_\alpha$ source (λ = 1.5406 Å)], scanning from 25° to 90°, with a 0.5° step size, incident angle. A glass holder was used to secure the sample. The XRD patterns in the films were identified with the support of ICDD-JCPDS data files [57]. The crystallite size of the CoFe was calculated using the Debye-Scherrer equation [58]. Additionally, the thickness of the thin film was measured using XRR analysis with a Cu K$_\alpha$ X-ray source (λ = 1.5406 Å). The same samples were used to simulate the specular XRR spectra using PANalytical X'pert software

*[X'Pert Reflectivity, v1.2a]*. Here, both GI-XRD and XRR measurements were conducted using the *PANalytical X'pert PRO* X-ray diffractometer. To obtain information about UMA, hysteresis curves were collected at different in-plane azimuthal angles (Φ) from samples annealed at different temperatures, using a homemade setup for the Longitudinal Magneto-Optical Kerr Effect (L-MOKE). The angular dependence of in-plane/parallel plane coercivity and squareness ratio was measured by rotating the samples at fixed angles. Furthermore, the magnetic properties of the films were further investigated in both in-plane and out-of-plane plane directions using a vibrating sample magnetometer (VSM) at RT *[Microsense EZ11 VSM]*. Interfacial oxidation states and chemical shifts in the as-dep. and annealed samples were investigated using X-ray Photoelectron Spectroscopy (XPS) to track the diffusion of Boron and Oxygen atoms and other accompanying structural variations during the annealing process. XPS analysis was conducted under a base pressure of $1.6\times10^{-7}$ mbar using a *PHI 5000 Versa Probe III* spectrometer, which employed a monochromatic source of Al K$_\alpha$ (1486.7 eV) irradiation with a pass energy of 55 eV. All the observed XPS spectra were calibrated using the standard reference energy of the C-1s peak at 284.76 eV. Following the measurements, the recorded data were analyzed using software of *CasaXPS software (version 2.3.25PR1.0)*. The damping parameters of the samples were determined using a Lock-in-based ferromagnetic resonance system *[NanOsc]*. Frequencies ranged up to 16 GHz, with an in-plane magnetic field strength from 0 to 5000 Oe in field sweep mode. The obtained FMR spectra were fitted with the derivative of the Lorentzian function to obtain resonance field ($H_r$) and linewidth ($\Delta H$) values.

## 3. RESULTS AND DISCUSSION

The schematic illustration of the Pd(5 nm)/CoFeB(10 nm)/Pd(3 nm)/Ta(10 nm) stacks is shown in Fig.1 (a). XRR analysis was employed for the as-dep. multilayer, as depicted in Fig.1 (b). In this case, the fitted XRR data closely matches the experimentally recorded data. The persistent presence of Kiessig fringes (XRR oscillations) up to 2θ = 6° indicates the films' high surface quality and their relationship with the interface. The simulated parameters for the as-deposited sample, including thickness, film density, and interface roughness, are presented in Table 1.

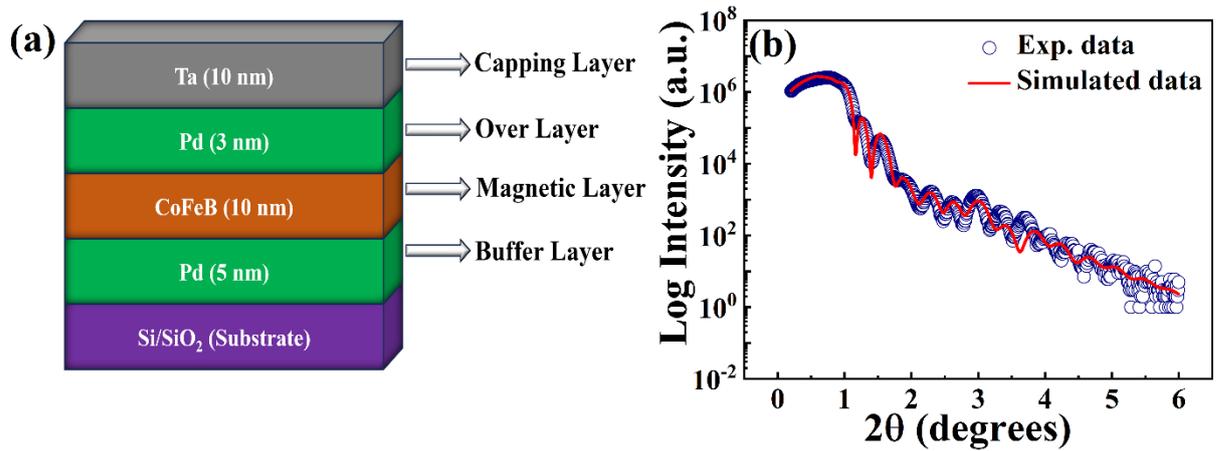

**Fig.1** (a) Schematic diagram of Pd/CoFeB/Pd/Ta multilayer films fabricated on Si/SiO$_2$ substrates at RT, (b) displays the fitted XRR data of the as-grown film.

**Table 1.** The simulated parameters for the thickness (t), density (ρ), and interface roughness (σ$_i$) of an as-dep. Si/SiO$_2$//Pd(5nm)/CoFeB(10nm)/Pd(3nm)/Ta(10nm) film. These parameters were determined through the fitting of experimentally obtained XRR profiles.

| S. No | Simulated parameters | Thin Film Layers | | | | |
|---|---|---|---|---|---|---|
| | | Pd | CoFeB | Pd | Ta | Ta$_2$O$_5$ |
| 1 | t (nm) | 5.2 ± 0.6 | 10.0 ± 0.2 | 3.1 ± 0.3 | 9.0 ± 0.3 | 2.0 ± 0.7 |
| 2 | ρ (g/cc) | 13.2 ± 0.4 | 7.1 ± 0.1 | 12.20 ± 0.02 | 16.61 ± 0.01 | 6.0 ± 1.0 |
| 3 | σ$_i$ (nm) | 0.414 ± 0.007 | 0.3 ± 0.3 | 0.3 ± 0.1 | 0.4 ± 0.2 | 0.5 ± 0.2 |

The density of each layer is determined to be either very close to or slightly smaller than their respective bulk values. The measured thickness of individual layers matches the nominal thickness values. Moreover, the overall roughness is identified to be less than 0.7 nm, indicating the excellent quality of these CoFeB films sputtered under high vacuum conditions.

### 3.1 Structural Analysis

To identify the crystalline nature of the multilayer films, we conducted GI-XRD measurements on as-deposited and annealed samples at 300ºC, 400ºC, 500ºC, 600ºC and 700ºC for Pd/CoFeB/Pd/Ta. The results are presented in Fig. 2. We observed two distinct diffraction patterns: one corresponding to Pd/PdO (111) and another to the substrate peak of Si (400). In the as-deposited film, there were no CoFeB diffraction peaks, indicating that CoFeB remained in an amorphous state. However, after annealing at 300ºC, we began to see diffraction peaks, suggesting a transformation from an amorphous to a partially nanocrystalline state for CoFeB [59]. It's important to note that high-intensity Bragg peaks were not observed, which is consistent with the lack of lattice coherence in these amorphous stacks and with previous reports on Co/α-$Al_2O_3$ and [CoFeB/MgO]$_{14}$ multilayers [60, 61].

Crystallization of CoFe embedded in the CoFeB amorphous matrix was observed in films annealed above 300°C, as reported by J. Sinha *et al.* [62]. With further increases in annealing temperatures, ranging from 400ºC to 700ºC, a small peak corresponding to bcc-CoFe (110) *[JCPDS card No. 00-044-1433]* became significantly visible at 2θ ≈ 45°. XRD also revealed the presence of $Ta_2O_5$ and PdO, which exhibited texturing in (110) and (111), respectively. These diffraction patterns appeared in the films annealed at 600ºC and 700ºC, indicating that higher annealing temperatures promoted the formation of oxides. Furthermore, the interdiffusion and oxidation at the interfaces of these samples will be discussed in the forthcoming section of the XPS analysis. In conclusion, we can deduce that, with increasing annealing temperature, the initially amorphous CoFeB film gradually transitions into a crystalline state.

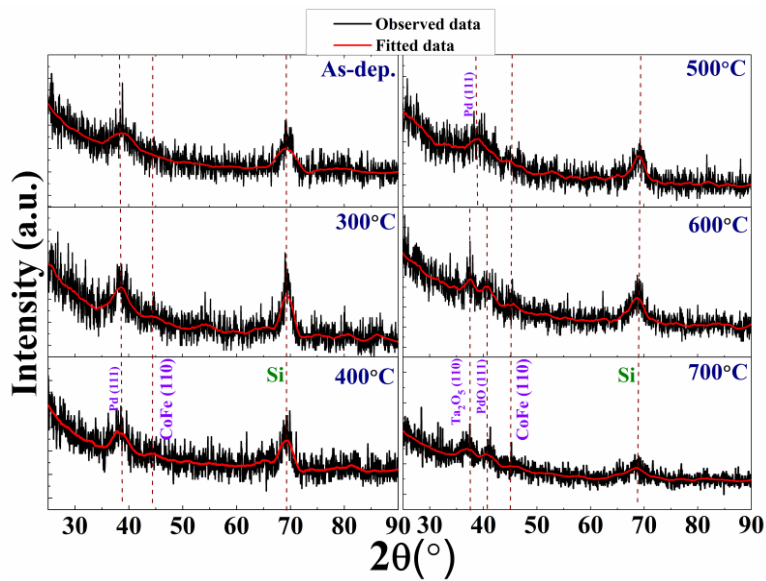

**Fig.2** X-ray diffraction patterns of Pd(5 nm)/CoFeB(10 nm)/Pd(3 nm)/Ta(10 nm) films deposited on a silicon dioxide substrate. The patterns are shown for various annealing temperatures, including as-dep., 300°C, 400°C, 500°C, 600°C and 700°C.

Based on the XRD data, we can determine the crystallite size of the CoFeB films using Debye Scherrer´s equation, as described in A.L. Patterson's work [58].

$$D = \frac{K\lambda}{\beta \cos\theta} \quad (1)$$

In this equation (1), D, K, β, and λ are the average crystallite size (nm), the crystallite shape factor constant, the full width at half maximum (FWHM) (radian), and the wavelength of the X-rays, respectively. The average crystallite size of the CoFeB sample is calculated to be ~ 22 nm, ~ 26 nm, ~ 27 nm, ~ 30 nm, and ~ 31 nm for the annealed 300°C, 400°C, 500°C, 600°C and 700°C films respectively.

## 3.2 Interfacial Oxidation/Diffusion

The XPS is a powerful technique that has been extensively utilized to investigate the impact of $T_A$ on the evolution of interfacial oxidation states and atom migration. The oxidation states at the interfaces are highly sensitive to thermal treatment owing to the diffusion and redistribution of oxygen atoms [63, 64]. On the other hand, interlayer diffusion and oxidation at the interfaces between Heavy metal and Ferromagnet layers play pivotal roles in affecting magnetic anisotropy and dynamic properties within the films. We performed XPS measurements on selected as-deposited and annealed (600ºC and 700ºC) Si/SiO$_2$//Pd(5nm)/CoFeB(10 nm)/Pd(3 nm)/Ta(10 nm) multilayers. Results confirm the presence of Pd, Co, Fe, B, Ta, O and C elements. The observed peak at 284.76 eV indicates the binding energy of C-1s and is considered a reference spectrum ascribed to the existence of carbon in the atmosphere.

The Core level XPS spectra of Co 2p and Fe 2p for the Pd(5 nm)/CoFeB(10 nm)/Pd(3 nm)/Ta(10 nm) multilayer film, both as-dep. and annealed at various temperatures, are shown in Fig. 3 (a)-(b). The peaks located at ≈ 778.3 eV and ≈ 793.3 eV are related to metallic Co $2p_{3/2}$ and $2p_{1/2}$ [65], while the peaks at ≈ 780.4 eV and ≈ 795.6 eV correspond to the metal oxide of CoO$_x$. Thus, both Co and O elements were detected in the three sets of samples: as-dep., at 600ºC, and 700ºC (Fig. 3 (a)). After Shirley's background correction, both Co $2p_{3/2}$ and Co $2p_{1/2}$ are in good agreement with the earlier reports [66, 67]. The Co 2p peak can be split into two main peaks at binding energies of ≈ 781.7 eV (Co$^{2+}$ $2p_{3/2}$) and ≈ 796.5 eV (Co$^{2+}$ $2p_{1/2}$), representing the strong presence of the Co-O bond for all the samples. The fitting of the Fe 2p spectra is shown in Fig. 3 (b). The peaks located at ≈ 706.7 eV and ≈ 720.4 eV are related to the metallic Fe $2p_{3/2}$ and Fe $2p_{1/2}$, while the peaks at ≈ 710.9 eV and ≈ 724.5 eV correspond to iron oxide $Fe_2O_3$. It is speculated that both metallic (Co and Fe) and metal oxides (Co-O and Fe-O) bonding coexist, with the oxidized forms of Co and Fe being more predominant in the annealed samples compared to the as-dep. multilayer film. Moreover, the presence of Co and Fe content in the CoFeB film is found to fluctuate with $T_A$, and traces of oxidation are detected in both Co and Fe.

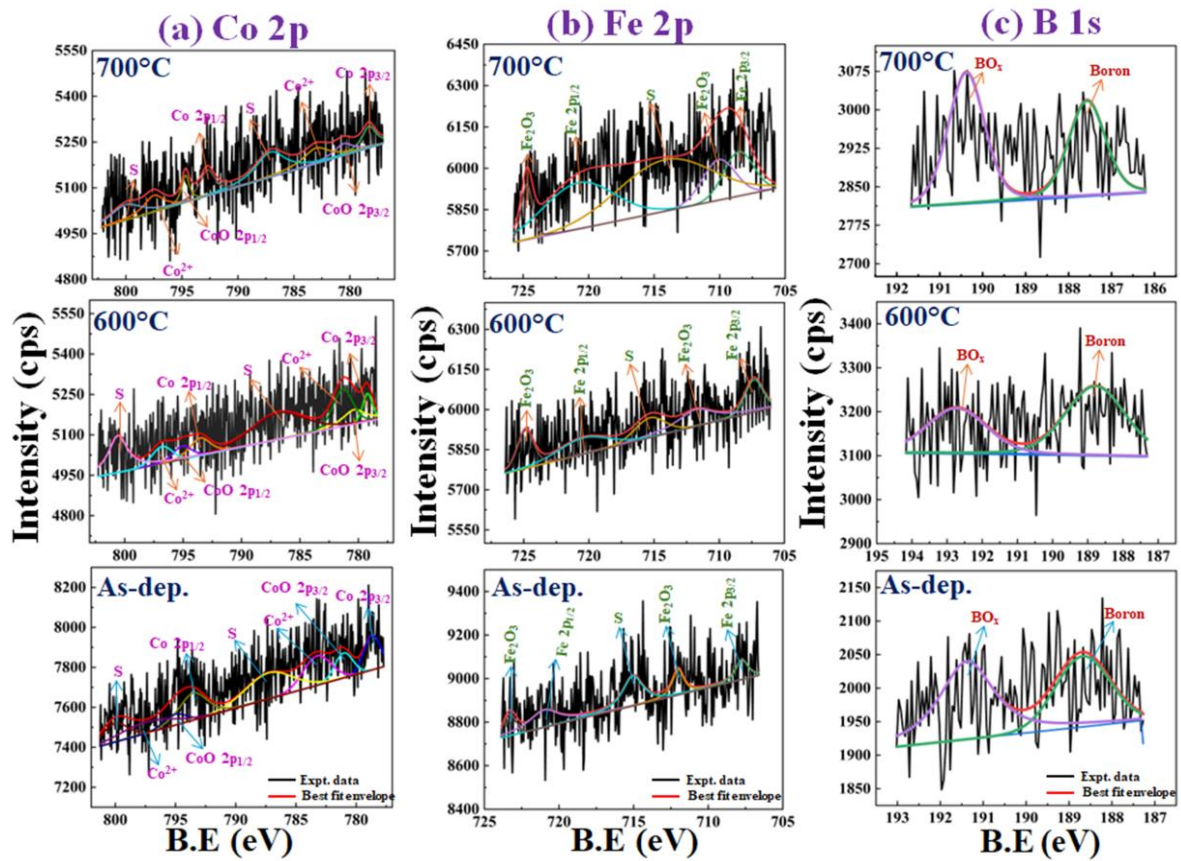

**Fig.3** XPS spectra of Co (a), Fe (b) and B (c) for both the as-dep. state and after annealing at 600ºC and 700ºC in stacks. The 'S' labels indicate the presence of satellite peaks.

In Fig. 3 (c), the core level spectra of B 1s are displayed both before and after annealing at various temperatures. Two distinct peaks in these spectra are attributed to two chemical states for B 1s: i) Boron originating from the CoFeB film, and ii) $BO_x$, associated with the more oxidized form of boron. These observations are consistent with previous reports [63, 68]. In the as-dep. sample, the B 1s spectra exhibit a strong signal at ≈188.7 eV, indicative of metallic boron. This signal reflects the covalent bond interaction between the metalloid and transition metals, such as B-CoFe. A signal of oxidized B 1s ($BO_x$) is detected in the as-dep. sample, with a binding energy of ≈191.4 eV. This oxide source is believed to naturally oxidize boron atoms in the vicinity of the interface [69, 70]. Following the annealing process, the metallic boron spectrum at approximately 188.8 eV (600°C) and 187.5 eV (700°C) significantly diminishes, while the boron oxide peak at around 192.7 eV (600°C) and 190.4 eV (700°C) intensifies unexpectedly. This behavior can be attributed to B's strong affinity for oxygen compared to Pd. During annealing, B migrates from the CoFeB layer to the adjacent Pd film, where it competes with Pd for oxygen (O) [47]. The recorded intensity of the B signal strength is as follows for the as-dep., 600ºC annealed and 700ºC annealed samples, respectively: 2.90 %, 2.61 %, and 1.48 %. Similar behavior has been observed in previously reported CoFeB/MgO/CoFeB/Ta stacks, where there is a transformation from the reduction of metallic boron to an increase in $BO_x$ following the annealing process [63, 69]. Furthermore, the observation of a wide boron oxide spectrum can be attributed to the formation of $B_2O_3$ and B suboxide contributions [69]. The reduction in boron intensity is due to the decrease in metallic boron from the CoFeB layer and the increase in the concentration of $BO_x$ due to the incorporation of oxygen atoms during annealing.

XPS spectra of the Pd element were obtained for both the as-dep. and annealed samples, as shown in Fig. 4 (a). Two orbitals, $3d_{5/2}$ and $3d_{3/2}$, corresponding to the Pd peaks were examined. In order to comprehensively analyze the interferences within the Pd/CoFeB/Pd films, depth analysis was conducted up to a depth of ~ 30 nm. In the spectra, metallic Pd ($Pd^0$) $3d_{5/2}$ and $3d_{3/2}$ peaks were observed at binding energies of ≈ 335.2 eV and ≈ 340.5 eV, respectively, showing a strong signal for the as-dep. sample, which is in good agreement with the literature [71-74]. Following annealing at 600ºC, the $Pd^0$ component significantly weakened, while the other component was identified as palladium oxide (PdO), present throughout the films. The peaks at ≈ 336.1 eV (Pd $3d_{5/2}$) and ≈ 341.9 eV (Pd $3d_{3/2}$) were associated with $Pd^{2+}$, indicating the presence of PdO. Additionally, a strong satellite peak (S1) for PdO ($Pd^{2+}$) with an energy separation of ≈ 2.6 eV was observed. The contributions of satellite peak S1 to the intensity of the Pd 3d spectrum were 2.48%. With a further increase in annealing temperature to 700ºC, an oxidic component became more pronounced in the Pd 3d XPS spectra. The contributions of the S1 and S2 peaks to the Pd 3d intensities were 2.27 % and 5.32 % respectively. The presence of S1 and S2 is attributed to Pd, and this could be related to a charge transfer process, although the exact nature of these satellites remains unclear [72, 74]. Based on the XPS spectra of the Pd investigation, it is inferred that a change in the oxidation level suggests the presence of a transition layer with an enhanced oxidation concentration at the interface between Pd and CoFeB. In our case, the metallic Pd component diminishes with the presence of the $BO_x$, indicating that the formation of Pd state oxidation (PdO) may result from the consumption of oxygen by boron.

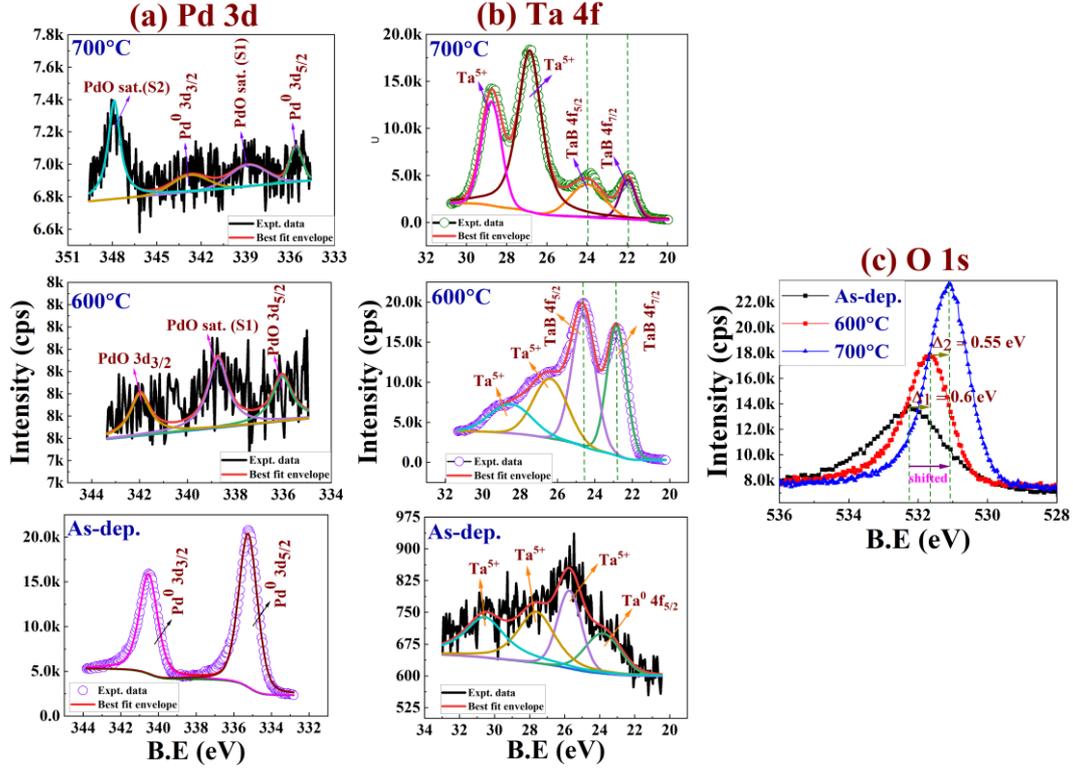

**Fig.4** XPS Core-level spectra for (a) Pd 3d, (b) Ta 4f and (c) O 1s, captured both before [as-dep.] and after annealing at 600 °C and 700 °C for Pd/CoFeB/Pd/Ta films.

We further conducted an in-depth analysis of the XPS spectra of Ta (Fig. 4 (b)). In the as-dep. sample, we observed oxidation peaks, alongside the metallic peak of $Ta^0$ $4f_{5/2}$ at ≈ 23.8 eV [75, 76]. This occurrence was expected because the samples were transferred from the sputtering unit to the XPS system, resulting in $TaO_x$ formation due to exposure to ambient air. Consequently, peaks corresponding to $Ta_2O_5$ (considered as $Ta^{5+}$) were observed in both the as-dep. and annealed films at 600°C and 700°C. Our findings align with Pauling's electronegativity scale for the elements, ranking Ta (1.5) < Fe (1.83) < Co (1.88) < B (2.04) < Pd (2.2) < O (3.44). This ranking supports our observation that Ta exhibits a higher affinity for oxygen than the other elements [51, 70]. Additionally, the presence of a thicker Ta capping layer effectively shielded the underlying Pd and CoFeB layers from oxidation at RT. In our study, we identified distinct Ta 4f states in the as-dep. sample, specifically $Ta^0$ $4f_{5/2}$ at ≈ 23.8 eV, $Ta_2O_5$ $4f_{7/2}$ at ≈ 25.7 eV, and $Ta_2O_5$ $4f_{5/2}$ at around 27.6 eV. Upon annealing at 600°C, we observed unexpected peaks at ≈ 24.6 eV and ≈ 22.8 eV, corresponding to the formation of TaB [53]. Notably, the intensity of these TaB peaks exceeded that of the $Ta_2O_5$ ($Ta^{5+}$) peaks, indicating that the Ta capping layer absorbed boron atoms from the underlying CoFeB layer during annealing. The observed shift in binding energy for the Ta oxide peaks after annealing at 600°C likely resulted from the formation of TaOB, with peaks at ≈ 26.4 eV and ≈ 28.5 eV [76]. In the 700°C annealed samples, we observed further significant shifts: the $Ta_2O_5$ $4f_{7/2}$ peak at ≈ 26.8 eV and $Ta_2O_5$ $4f_{5/2}$ peak at ≈ 28.7 eV, alongside lower intensity peaks for TaB $4f_{7/2}$ (≈ 22.0 eV) and TaB $4f_{5/2}$ (≈ 24.0 eV). The notable increase in $Ta_2O_5$ peak intensity in the 700°C sample suggests a slight thickening of the tantalum oxide layer due to annealing and exposure to the atmosphere. The optimal annealing temperature facilitated boron migration from the CoFeB layer to the Ta layer, forming TaB. Additionally, TaOB may have formed as boron atoms migrated into the adjacent $Ta_2O_5$ (Ta-O-Ta) layer of the Ta capping layer, resulting in the TaO-B bond [53,76-78]. The formation of TaB and TaOB induces compositional changes at the interfaces, which can influence critical magnetic properties dependent on interface sharpness. Transition metal borides, including Co-B, Fe-B, and Ta-B, exhibit enthalpy formation values of -32 kJ/mol, -36 kJ/mol, and -78 kJ/mol, respectively [79, 80]. These data highlight that Ta-B has the most negative enthalpy formation value, indicating its superior thermodynamic stability. Consequently, the intensified Boride peak likely reflects the dominance of the Ta-Boride layer over the CoFe-Boride layer, consistent with previous findings [70, 72].

The interfacial oxidation is highly sensitive to the annealing temperature due to the migration or redistribution of oxygen atoms [66]. In Fig 4 (c), the O 1s spectrum of the Pd/CoFeB/Pd/Ta multilayer films is presented for both as-dep. and annealed samples at 600ºC and 700ºC. All the samples exhibit a single spectrum, with the as-dep. sample's O 1s peak is located at ≈ 532.2 eV. Notably, a significant chemical shift (Δ ≈ 0.6 eV) towards lower binding energy is observed in the O 1s peaks for the selected 600ºC annealed stack compared to the as-dep. sample. Upon further annealing at 700ºC, the corresponding O 1s peak shifts to 531.05 eV. This demonstrates the crucial role of oxygen atoms in influencing interfacial interactions. The presence of chemical shifts ($\Delta_1$ and $\Delta_2$) in the annealed samples can be attributed to the reduction in oxygen vacancy density. This reduction may be a result of the interfacial oxygen diffusion through redox reactions (or) the redistribution of oxygen atoms within the Pd/CoFeB/Pd/Ta layers during annealing. Moreover, after the annealing process, the release of oxygen atoms decreases, leading to the formation of more Co, Fe, B, Pd, and Ta oxides. Consequently, the $TaO_x$ and $BO_x$ spectra remain stable up to $T_A$ of 600ºC.

Conversely, TaB formation and an excess of oxygen migration into the stacks are observed in the 700ºC sample. Generally, TaB becomes mixed with the metallic Ta (Ta$^0$) layer during adequate annealing temperatures [70]. In summary, the variations in the TaO$_x$ and BO$_x$ spectra, along with changes in the Co 2p, Fe 2p, and Pd 3d spectra, indicate distinct oxygen diffusion behaviors in the 600ºC and 700ºC samples. In the sample annealed at 700ºC, thermally activated oxygen atoms strongly diffuse into the Ta, Pd, and CoFeB layers.

### 3.3 Magnetic Anisotropy Properties

*3.3.1 In-plane Magnetic anisotropy dependence on annealing*

The magnetic anisotropy in the Pd/CoFeB/Pd/Ta multilayer at various annealing temperatures was investigated by analyzing the angle-dependent hysteresis loops obtained by Magneto-Optical Kerr Effect (MOKE) with a magnetic field applied in the in-plane direction. Longitudinal angle-dependent MOKE hysteresis curves were obtained at RT, ranging from 0° to 360° with a step size of 10°, for all the as-dep. and annealed samples. During annealing, structural evolution generally involves two critical stages that significantly impact magnetic anisotropy: 1) stress release and 2) crystallization. Stress-induced anisotropy exhibits uniaxial symmetry, while crystallization-induced anisotropy shows cubic symmetry. The dominance of cubic anisotropy in the film annealed at 700°C is attributed to the development of CoFe nanocrystallites. These nanocrystallites weaken the exchange interaction among CoFe grains, thereby enhancing the local average magnetocrystalline anisotropy with cubic symmetry [81].

Fig. 5 (a)-(f) displays the in-plane normalized M-H (L-MOKE) loops of the as-dep. and annealed Pd/CoFeB/Pd/Ta multilayers along the Easy Axis (EA) and Hard Axis (HA). These measurements revealed a strong magnetic anisotropy presence up to 600°C. In the initial stages of crystallization for the 300°C and 400°C annealed samples (depicted in Fig. 5 (b)-(c)), the hysteresis loops exhibit different shapes along the two orientations, clearly indicating the presence of uniaxial magnetic anisotropy. The in-plane anisotropy field ($H_{k//\ [MOKE]}$) is observed to be ~ 36 Oe and ~ 48 Oe for the 300°C and 400°C samples, respectively (as shown in Fig. 5 (h)). Generally, the $H_{k//}$ is determined by the difference between the saturation field of the in-plane hysteresis loop evaluated along the hard and easy axes. These values are relatively higher than those of the as-dep. film (~ 28 Oe) and are possibly associated with the absence of internal stress relief in the sample as crystallization progresses [82]. Further, the decrease in the $H_{k//\ [MOKE]}$ values (~ 25 Oe and ~ 18 Oe for the 500°C and 600°C multilayers, respectively) observed with increasing $T_A$ up to 600°C is attributed to the release of substrate-induced strain at specific temperatures [82].

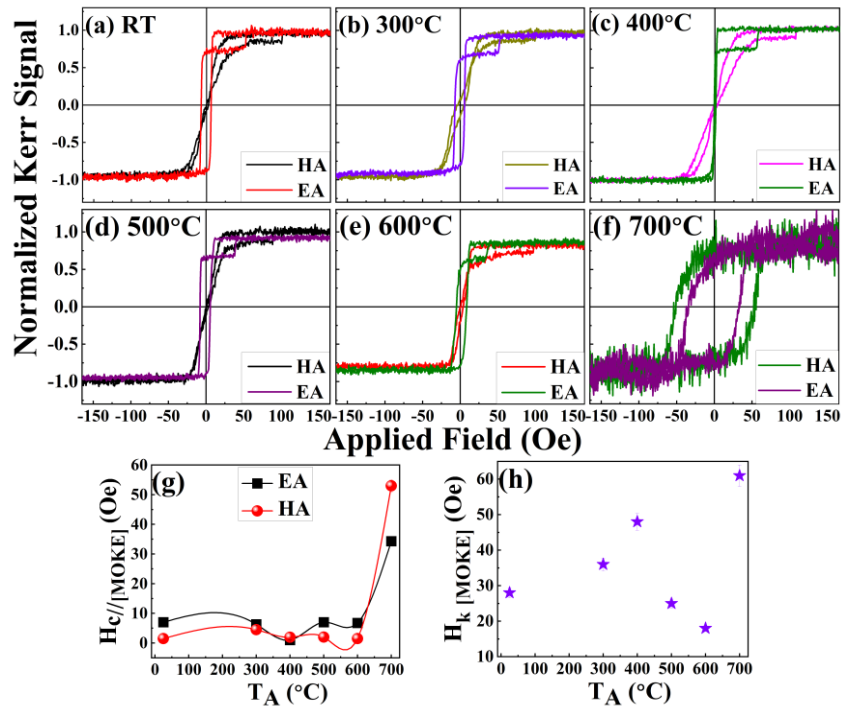

**Fig.5** (a)-(f) L-MOKE hysteresis loops of multilayer annealed from RT to 700°C with the magnetic field applied along both the easy and hard axes. (g) the in-plane coercivity ($H_{c//[MOKE]}$) measured as a function of $T_A$. (h) Variation of anisotropy field $H_{k// [MOKE]}$ field versus $T_A$.

The hysteresis loops measured with a magnetic field applied along the easy axis exhibit a squared shape with small coercivities from the as-dep. state to 600°C, which is typical for easy magnetization switching, either by domain reversal (domain nucleation) or domain wall motion. In the current study, magnetic anisotropy (UMA) up to 600°C remains stable even after crystallization. The presence of Boron in the CoFeB may reduce coercivity and promote the in-plane UMA [83, 84]. A UMA dominated by the evolution of effective cubic anisotropy occurs during the complete transformation from the amorphous to the crystalline state of CoFe at elevated temperatures, as supported by our XRD analysis. As the CoFe nanocrystallites evolve, exchange coupling weakens among the CoFe grains, consequently enhancing the local average cubic magneto-crystalline anisotropy. Further annealing at 700°C resulted in an abrupt change in the HA hysteresis loop evidenced by a rise of the remanent magnetization. As shown in Fig. 5(f), the in-plane coercivity ($H_{c//[MOKE]}$) of both the EA and HA loops substantially increased, which could be attributed to:

1. The presence of domain wall pinning at grain boundaries [85].
2. The migration of Boron from the CoFeB layer to the adjacent Pd or even Ta layer favors the crystallization of CoFe. CoFe films typically demonstrate higher coercivity than CoFeB films [86, 87]. Additionally, even the presence of a small amount of hard magnetic boride phase could enhance coercivity [88, 89].
3. The oxidation of the CoFeB film with the formation of CoO, FeO and $BO_x$ and the occurrence of atomic interdiffusion of TaB, B, and O could also impact the magnetic properties [70, 86, 90]. The formation of metal oxides has been extensively discussed within the context of our XPS results (in Fig. 3 and 4).

Moreover, earlier reports have shown that $H_{c//}$ depends on fabrication parameters and other notable structural factors such as deposition rate, sputtering bias voltage crystallite size, pinning holes, film, and more. Additionally, local structural properties such as defects, changes in strain, and grain sizes can also affect both in-plane coercivity and squareness ratio [91].

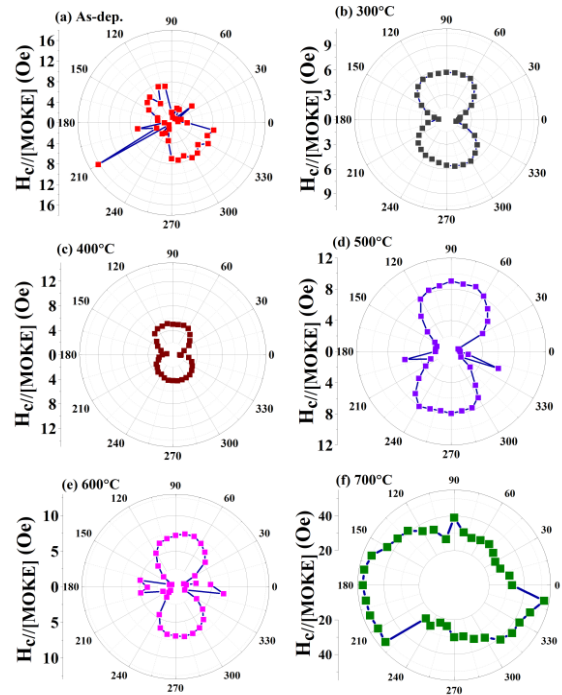

**Fig.6** Polar plot showing the variation of $H_{c//[MOKE]}$ at different $T_A$; (a) as-dep., b) 300°C, (c) 400°C, (d) 500°C, (e) 600°C and (f) 700°C, as a function of the azimuthal angle variation.

Fig. 6 presents plots of in-plane $H_{c//[MOKE]}$ as a function of Φ for both as-dep and annealed samples obtained from angle-dependent MOKE measurements. Additional characterization of $M_r/M_{s//}$ can be found in Fig.S1 (Supplementary Material). In the as-dep. film, the angular evolution of $H_{c//[MOKE]}$ exhibits a relatively weak two-fold symmetry. Upon annealing at 300°C and 400°C, the magnetic anisotropy displays two-fold symmetry with dumbbell-like characteristics, as shown in Fig. 6 (b) and (c). Similar phenomena were previously observed in CoFeB films [89, 92]. Furthermore, annealing at 500°C and 600°C, two symmetric spikes form along the hard axis of the two-fold symmetry, indicating the presence of a weak higher-order contribution to magnetic anisotropy. This weak contribution arises from the onset of crystallization of the amorphous CoFeB layer and contributes to the uniaxial magnetic anisotropy [39]. This suggests that, in terms of the magnetic exchange length, the grain size created at these temperatures is not sufficiently large to generate a large coercivity [85]. Further annealing at 700°C leads to a significant increase of $H_{c//[MOKE]}$ and an isotropic magnetic anisotropy as shown in Fig. 6 (f). This could be due to the occurrence of almost complete oxygen diffusion and redox reactions within the multilayer [63, 93, 94] and it is consistent with our experimental XPS results. Further discussion of magnetic anisotropy is provided in the Supplementary Material.

### 3.3.2. Static magnetic characterization for in-plane and out-of-plane configuration

Static magnetic properties were studied, using a VSM at RT, to identify the type of magnetic anisotropy. Fig. 7 (a)-(f) displays the normalized in-plane and out-of-plane hysteresis loops of the Pd(5 nm)/CoFeB(10 nm)/Pd(3 nm)/Ta(10 nm) stacks annealed at different temperatures. In each of the M-H curves, saturation is achieved at 500 Oe when the magnetic field is parallel to the film plane. However, when the magnetic field is applied perpendicular to the plane, the hysteresis loops do not reach full saturation at 0.5 Tesla. This observation suggests that the easy axis of magnetization is within the plane of the film for all the samples. The annealing process can alter the targeted magnetic properties and prompt the rearrangement of atoms at their interfaces, effectively eliminating the disorder generated during the sputtering growth process [95]. The $H_{k// [VSM]}$ values, obtained by VSM, are shown in Table 2. The positive and negative signs of $H_{k\perp [VSM]}$ and $H_{k// [VSM]}$ represent the presence of PMA and IPA, respectively, although all our samples exhibit IPA properties. The stronger IPA exhibited by the as-dep. film, compared to the annealed samples, is attributed to the enhanced demagnetization energy. Moreover, as the $T_A$ rises from 600ºC to 700ºC, a relatively weaker IPA is observed in the sample annealed at 700ºC. This anisotropy shift may arise from a decrease in the demagnetization energy contributing to IPA and a minor enhancement in the magneto-crystalline anisotropy energy at their interfaces [96]. Consequently, significant interfacial changes become more pronounced in the 700ºC film.

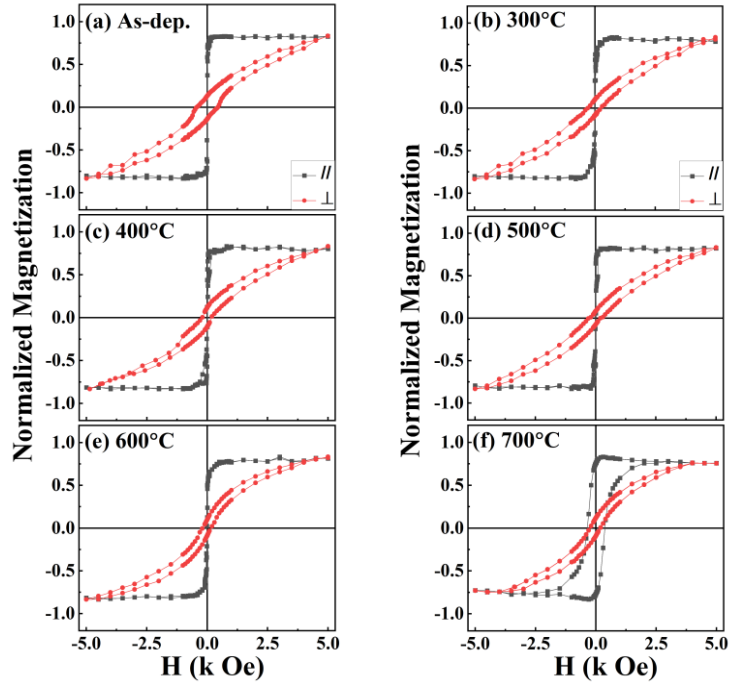

**Fig.7** (a)-(f) M-H loops measured at RT in both in-plane (in black) and out-of-plane (in red) directions corresponding to the Pd/CoFeB/Pd/Ta systems annealed up to 700ºC.

**Table 2:** The magnetic parameters; $M_{s//}$, $H_{c//}$, $H_{c\perp}$, $H_{k//}$, and $K_{eff}$ derived from VSM M-H loops for Pd/CoFe/Pd/Ta multilayers annealed at different $T_A$.

| $T_A$ (ºC) | Magnetic Properties | | | | |
|---|---|---|---|---|---|
| | $M_{s//}$ (emu/cc) | $H_{c//}$ (Oe) | $H_{c\perp}$ (Oe) | $H_{k//}$ (Oe) | $K_{eff}$ ($\times 10^6$ erg/cc) |
| As-dep. | 1165 ± 47 | 6 ± 1 | 428 ± 17 | -4851 ± 194 | -2.8 ± 0.1 |
| 300 | 1211 ± 49 | 15 ± 2 | 283 ± 11 | -4523 ± 181 | -2.7 ± 0.1 |
| 400 | 1340 ± 54 | 14 ± 2 | 198 ± 8 | -4516 ± 180 | -3.0 ± 0.1 |
| 500 | 1482 ± 59 | 9 ± 1 | 252 ± 10 | -4500 ± 178 | -3.3 ± 0.1 |
| 600 | 1276 ± 51 | 4 ± 1 | 175 ± 7 | -4346 ± 174 | -2.7 ± 0.1 |
| 700 | 1112 ± 44 | 366 ± 15 | 216 ± 9 | -3894 ± 156 | -2.1 ± 0.1 |

The impact of $T_A$ on $M_{s//}$, $H_{c//[VSM]}$, and $H_{c\perp[VSM]}$ is summarized in Fig. 8 (a) and (b). The $M_{s//}$ of the CoFeB samples gradually increases from 1165 ± 47 emu/cc to 1482 ± 59 emu/cc as the $T_A$ rises from RT to 500ºC, as shown in Fig. 8 (a). This increase in $M_{s//}$ can be attributed to the reduction of Boron atoms (magnetic impurities) within the amorphous structure of the CoFeB film during annealing [97]. According to the rigid band model, the magnetic moment of the CoFe (transition metals) increases as the Boron (metalloid) concentration decreases. This phenomenon occurs due to the significantly smaller atomic size of Boron compared to other atoms such as Pd, Co, Fe, and Ta. Moreover, Boron diffusion is simpler and faster than that of other elements, leading to changes in magnetization during the annealing process [97]. Furthermore, as the annealing temperature increases from 500ºC to 700ºC, the $M_{s//}$ drastically reduces to 1112 ± 44 emu/cc. This reduction of $M_{s//}$ is probably attributable to the decrease of CoFeB thickness resulting from the intermixing of CoFeB and buffer (Pd) layer [98] and to the deepening oxidation at the interfaces [54] of Pd/CoFeB and CoFeB/Pd/Ta. Therefore, it is accountable that oxygen can diffuse into the Pd, Ta and CoFeB layers during higher annealing temperatures of 600ºC and 700ºC. Moreover, it is well established that the hybridization of B 1s with the CoFe 3d states leads to a reduction in the magnetic moments of Co or Fe and generates a small negative moment in the boron atoms [99, 100]. Our results show that magnetization, particularly the $M_{s//}$ values, is significantly influenced by the presence of CoO and FeO, as well as by oxidation occurring at interfaces such as PdO and $TaO_x$ layers. The detected oxidations are confirmed by the XPS data presented in Fig. 3 and 4 and XRD data in Fig. 2. The effect of Oxygen diffusion and oxidation mechanism on magnetization is in good agreement with recent reports [51, 63, 93].

Fig. 8 (b) displays the $H_{c//[VSM]}$ and $H_{c\perp[VSM]}$ for all the films at various temperatures. The in-plane coercivity ($H_{c//[VSM]}$) values for the as-dep. samples and samples annealed up to 600ºC range from approximately 4 Oe to around 15 Oe. In contrast, the sample annealed at 700ºC displays a significantly higher coercivity of approximately 366 Oe, as shown in Fig. 7 (f). The reason for the sudden increase in $H_{c//}$ in the sample annealed at a temperature of 700°C has been previously discussed. As observed in Fig. 8 (b), $H_{c\perp[VSM]}$ decreases as $T_A$ increases with a noticeable oscillation [$H_{c\perp[VSM]}$ decreases from 428 ± 17 Oe to 175 ± 7 Oe, as $T_A$ increases from RT to 700ºC]. The decrease in $H_{c\perp[VSM]}$ can be attributed to reduced domain wall pinning and/or nucleation, a higher degree of oxidation, and intermixing of heavy metals (in our case Pd and Ta) and CoFeB interfaces in the multilayer films [51, 63]. The crystallization of CoFe and oxidation at their interfaces contribute to the observed values of $M_s$ and $H_{c[VSM]}$.

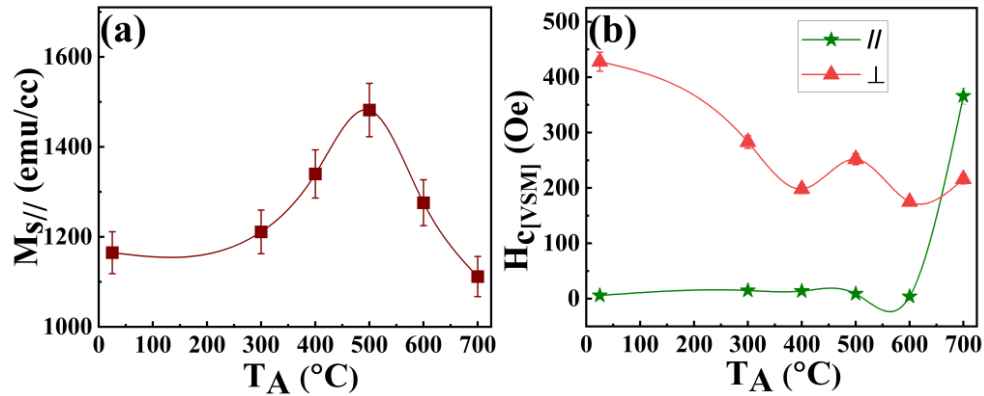

**Fig.8** Post-annealing temperature dependence of (a) $M_{s//}$ and (b) $H_{c//[VSM]}$ and $H_{c\perp[VSM]}$ for the Pd(5 nm)/CoFeB(10 nm)/Pd(3 nm)/Ta(10 nm) multilayer, spanning a temperature range from RT to 700ºC.

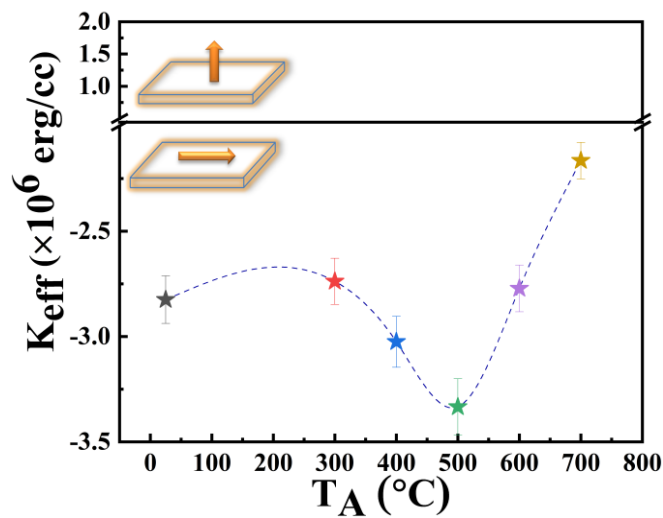

**Fig.9** $K_{eff}$ as a function of $T_A$ for Pd(5)/CoFeB(10)/Pd(3)/Ta(10) stacks.

To further understand the interfacial in-plane magnetic anisotropy of CoFeB films, we investigate the dependence of $K_{eff}$ on the different $T_A$ for Pd/CoFeB/Pd/Ta layers. The $K_{eff}$ equations are provided in the Supplementary Material. Fig. 9 illustrates the $T_A$ dependence of $K_{eff}$ for the Pd(5 nm)/CoFeB(10 nm)/Pd(3 nm)/Ta(10 nm) stacks. The maximum negative $K_{eff}$ value is estimated to be -3.33 ± 0.133 ×10$^6$ erg/cc for the film annealed at 500ºC. However, the smallest negative $K_{eff}$ value of -2.16 ± 0.086 × 10$^6$ erg/cc is observed for the 700ºC sample, showing a 42.5 % difference in $K_{eff}$ value compared to the 500ºC stack. Consequently, $K_{eff}$ negatively increases as the $T_A$ increases from RT to 500°C. This could be attributed to the development of enhanced demagnetization energy in the films. Nevertheless, further increasing the $T_A$ from 500°C to 700°C gradually decreases (negatively) the $K_{eff}$ of the multilayer. The decrease of $K_{eff}$ may be related to a larger degree of strain homogeneity of the CoFeB film [101, 102]. To conclude, the $T_A$ is a critical parameter for manipulating the magnetic properties of Pd/CoFeB/Pd/Ta multilayer films. Therefore, based on these findings, a high IPA energy is recorded for the 500°C sample, while a lower IPA energy is observed for the 700°C sample.

### 3.4 Magnetization Dynamics

The precession and relaxation of magnetization are commonly described by the Landau-Lifshitz-Gilbert formula, incorporating a phenomenological parameter known as Gilbert damping ($\alpha$) [103]. The $\alpha$ values are significantly influenced by factors such as surface or interface properties, adjacent layers, thermal treatment, microstructures, and the incorporation of dopants into the ferromagnetic layer [103-108]. Tuning of $\alpha$ in a dedicated device remains challenging, with a commonly employed approach being the post-annealing of the deposited film [104, 109]. This process agglomerates small islands into larger ones, improving their crystalline nature. It has been demonstrated that annealing temperature induces the partial crystallization of CoFe in layers like HM/CoFeB/MgO or CoFeB/MgO/CoFeB due to Boron diffusion [110-112]. The annealing temperature necessary to induce partial crystallization is determined by the boron content of the CoFeB layer [113]. Studies have reported decreased $\alpha$, coercivity, and effective magnetization with the formation of partial crystallization [114]. Low $\alpha$ in CoFeB is associated with partial crystallization induced by annealing [115], and the release of stress under annealing [81].

In-plane FMR measurement was conducted at RT to investigate the magnetization dynamics and Gilbert damping behavior of Pd/CoFeB/Pd/Ta multilayer films annealed at various temperatures. The analysis was performed in the field sweep mode, wherein a DC magnetic field was externally applied along the in-plane easy axis of the film stacks. FMR analysis was carried out over a frequency (f) range from 5 to 16 GHz along the in-plane easy axis. Fig. 10 (a) displays the derivative FMR absorption signal for the as-dep. sample. All FMR signals obtained at different f values were fitted using a derivative combination of symmetric and anti-symmetric Lorentzian functions to determine the linewidth ($\Delta H$) and resonance field ($H_r$) [refer Eqn. (9) in the Supplementary Material]. For comparison and better comprehension,

FMR spectra for all the samples were recorded at a fixed frequency of 9 GHz, centered at 0 Oe (depicted in Fig. 10 (b)). Annealing the as-dep. sample at 300ºC leads to a notable broadening of the linewidth. This broadening could be attributed to the crystallization process and the removal of stress in the samples through annealing, making them pinning-free [116]. The above-observed linewidth broadening behavior aligns well with findings from previous reports [115, 117, 118]. The frequency dependence of $H_r$ for the different annealed samples is fitted using Kittel's equation [119], providing details about the effective magnetization field ($4\pi M_{eff}$) and effective anisotropy field ($H_{k//}$),

$$f = \frac{\gamma}{2\pi}\sqrt{(H_r + H_{k//})(H_r + H_{k//} + 4\pi M_{eff})} \qquad (2)$$

Where, γ represents the gyromagnetic ratio ($1.856 \times 10^{11}$ Hz/T) and $H_{k//}$ is the in-plane uniaxial magnetic anisotropy field, typically less than $4\pi M_{eff}$. The value of $4\pi M_{eff}$ is associated with saturation magnetization ($4\pi M_S$) and interfacial anisotropy properties, as given by the equation [56, 120]:

$$4\pi M_{eff} = 4\pi M_S - \frac{2K_S}{M_S t_{CFB}} \qquad (3)$$

Here, $K_S$ and $t_{CFB}$ are the surface/interface anisotropy constant and CoFeB film thickness, respectively.

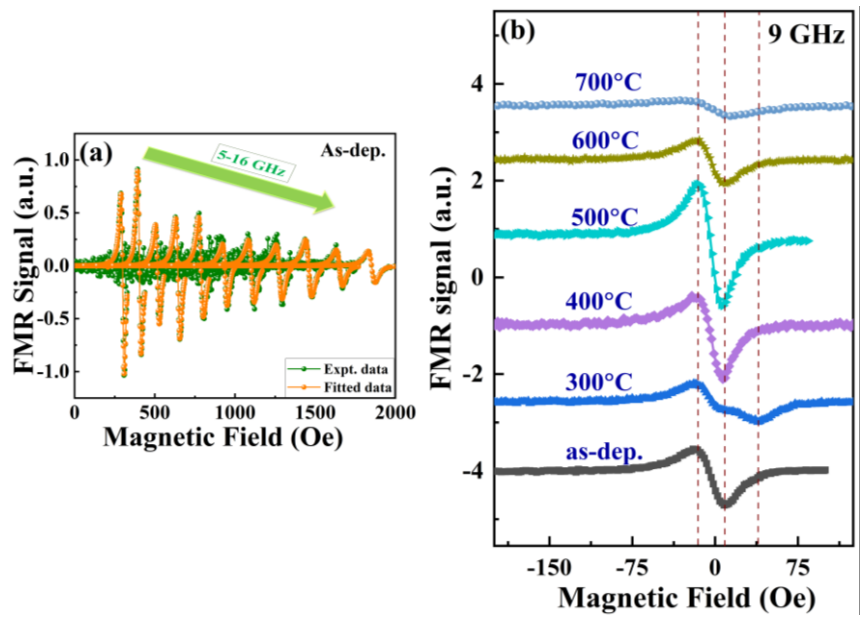

**Fig.10** (a) In-plane FMR signal for the as-dep. multilayer structure across a frequency range of 5-16 GHz, (b) FMR spectra recorded at 9 GHz along the easy axis of the samples, varying $T_A$.

Fig. 11 (a) shows the frequency dependence of $H_r$. Annealed samples exhibit higher $H_r$ values than the as-dep. sample, indicating that the magnetic field required to excite resonance is altered by post-annealing. O'Dell *et al* 2018 [116] reported that the annealing process induces a shift in $H_r$ due to the onset of partial crystallization of CoFe(B). The observed variation in $H_r$ towards higher fields as a function of $T_A$ suggests the potential presence of the CoFe intermediate state, which may manifest as an increase in the CoFe peak's intensity. To comprehend the impact of $T_A$ on the magnetization relaxation of the multilayer films, we investigate the dependence of $\Delta H$ as a function of f (see Fig. 11 (b)). $\Delta H_0$ and α parameters of the CoFeB film are extracted from the equation (4) [86, 106, 120],

$$\Delta H = \Delta H_0 + \Delta H_G + \Delta H_{TMS} \tag{4}$$

Here, $\Delta H_G = \frac{2\pi f}{\gamma}\alpha$. The first term in Equation (4) is the frequency-independent linewidth contribution known as inhomogeneous broadening ($\Delta H_0$), which usually arises due to magnetic inhomogeneity. The second term indicates the frequency-dependent magnetization-relaxation mechanism, where α is the parameter that describes the magnetization relaxation rate, whose value is essential for eligibility in STT/SOT/magnetic skyrmion-based MRAM device applications. The determination of the intrinsic part, α, representing the ability to convert precessional energy into heat through spin-orbit interaction, can be calculated from $\Delta H$ and scales linearly in frequency (f) [86, 87, 121]. The last term, $\Delta H_{TMS}$ represents the linewidth contribution owing to two-magnon scattering (TMS). The extrinsic part takes into account TMS. The TMS is a non-Gilbert-type damping contribution in which a uniform magnon of a precessional macrospin (FMR mode) scatters into a degenerate nonuniform short-wavelength magnon induced by material imperfections. In our system, contributions from magnon-phonon scattering, eddy current, and spin pumping effects are relatively weak, as evidenced by the linear dependence of $\Delta H$ vs f observed for all samples [103].

In Fig. 11 (c), $\Delta H_0$ is presented as a function of $T_A$. The plot illustrates a rapid increase in $\Delta H_0$ at 300ºC, followed by a drastic decrease to $T_A$ = 500ºC. Subsequently, the $\Delta H_0$ value slightly increased from $T_A$ = 500ºC to 700ºC. The $\Delta H_0$ value at $T_A$ = 300ºC is attributed to increased inhomogeneity due to the onset of crystallization, related to magnetic disorder formed by the interfacial roughness of the CoFeB interface. The observed decrease in $\Delta H_0$ values for the 400ºC and 500ºC annealed samples could result from stress release, leading to increases in UMA. The presence of oxidation (PdO and TaO$_x$) in the films indicates the migration of oxygen and redox reactions within the multilayer, which could significantly diminish the in-plane UMA nature. Hence, no traces of oxidation are detected in the diffraction peaks from RT to samples annealed at 500°C, as confirmed by our XRD data [101, 122]. The difference in UMA from the MOKE measurement in Fig. 5 and Fig. 6 may also indicate the linewidth decrease. A slight enhancement in ΔH$_0$ values in the 700ºC sample might be attributed to an increase of Boron segregations and/or agglomeration of Boron towards their interface, the formation of cubic anisotropy, and an increase in grain size, thereby contributing to an increasing linewidth [104, 122].

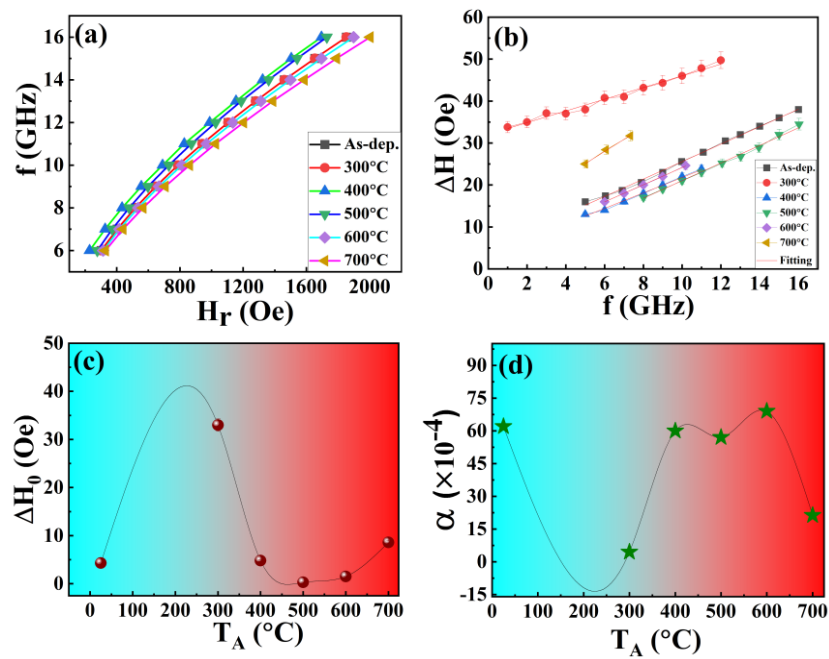

**Fig.11** (a) Frequency (f) vs resonance field (H$_r$) for all multilayer samples, (b) variation of ΔH as a function of f, (c) the dependence of ΔH$_0$ on T$_A$, and (d) the variation of α on T$_A$ (d). The lines are plotted merely to serve as a visual guide.

Fig. 11 (d) illustrates the dependence of α on T$_A$. Initially, the damping decreases from as-dep. to 300ºC, and then increases up to T$_A$ ≤ 600ºC. The initial reduction in α from 0.0062 to 0.00045 for T$_A$ ≤ 300ºC can be attributed to variations in the structural phase of CoFeB, particularly to the decrease of surface/interface inhomogeneities, the release of stress under annealing, partial crystallization, and consider the possibility of an intermediate state existing between the amorphous and nanocrystalline phases [122]. There is a rapid increase of α for the samples annealed at 400ºC, 500ºC, and 600ºC, with α values of 0.0067, 0.0057, and 0.0069, respectively. This increase may be attributed to the modifications of the interface between CoFeB, Pd, and Ta during annealing, leading to magnetic disorder, larger surface roughness, and the relaxation of uniaxial anisotropy. Upon complete crystallization, the sample annealed at 700ºC exhibits oxidation, inter-diffusion, the presence of even a small amount of hard magnetic boride phase and an enhanced crystalline-induced cubic magnetic anisotropy, resulting in a decrease in the damping constant [81, 122].

# 4 Conclusions

Our investigation methodically explores the intricate effects of annealing temperature on the structural, magnetic, and interfacial oxidation/diffusion properties of Si/SiO$_2$//Pd(5 nm)/CoFeB(10 nm)/Pd(3 nm)/Ta(10 nm) multilayer films, uncovering correlations among these properties. Annealing has a profound impact on the uniaxial magnetic anisotropy (UMA) energy, specifically, films annealed at 500°C exhibit an enhanced in-plane UMA, achieving a maximum K$_{eff}$ of ≈ 3.33 × 10$^6$ erg/cc, attributed to interlayer atomic migration and the onset of over-oxidation at the Pd/CoFeB/Pd/Ta interfaces. This enhancement coincides with the increasing instability and oxidation of Co and Fe in the CoFeB layer as the T$_A$ increases, notably leading to the oxidation of metallic B and the reduction of the metallic Pd component, indicative of forming a PdO state. The oxidation phases of Pd and Ta intensify above 500°C; though, TaO$_x$ and BO$_x$ remain stable, maintaining UMA up to a T$_A$ of 600°C. Further annealing at 700°C promotes the formation and the diffusion of TaB and TaOB states, increasing the interface width and decreasing the thickness of Ta and CoFeB. Consequently, this leads to modifications in magnetic anisotropies, namely, degradation of the in-plane magnetic anisotropy. Hence, at a temperature of 700°C, the reduced release of oxygen atoms leads to an increased formation of Co, Fe, B, Pd, and Ta oxides, resulting in a deterioration of UMA due to excessive oxygen diffusion into the stacks. Moreover, magnetic reversal is significantly changed by the interaction between uniaxial and cubic anisotropies. These findings underscore the complex interplay between thermal treatments and the resulting structural and chemical changes within multilayer films, significantly impacting their magnetic properties. The T$_A$ also plays a crucial role in tuning the magnetic damping properties of the material, an ultra-low damping constant (α) of 4.5 ± 0.15 × 10$^{-4}$ was observed after annealing at 300°C, which is attributed to the partial crystallization of CoFe, a decrease of surface/interface inhomogeneities and the release of stress under annealing. Conversely, a decrease in α at a higher temperature (700ºC) can be attributed to a reduced impact of structural disorder and an excess of oxygen migration into the multilayer stacks. The observed low α value in the sample at 300°C together with high thermal stability (large $K_{eff}$ value) up to 500°C with in-plane UMA (two-fold symmetry), offers significant advantages for Spin Torque (STT/SOT)-based MRAM applications. Moreover, this investigation provides crucial insights into optimizing annealing processes for enhancing the performance of materials in advanced spintronics applications.

## Acknowledgments

L. Saravanan acknowledges to FONDECYT Postdoctorado 2022 ANID, 3220373. C. Garcia acknowledges the financial support received by ANID FONDECYT/Regular 1201102, ANID FONDECYT/Regular 1241918, ANID FONDEQUIP EQM140161, and ANID FONDEQUIP EQM 150094. This work was also supported by the European Union's Horizon 2020 research and innovation program under the Marie Sklodowska-Curie Grant Agreement No. 734801 (MAGNAMED) and No. 101007825 (ULTIMATE-I).

**Highlights**

- Annealing temperature ($T_A$) has a significant impact on the structural, interfacial oxidation, and static and dynamic magnetization properties of Pd/CoFeB/Pd/Ta structures.

- A low damping value of $4.5 \times 10^{-4}$ is achieved for the annealed film.

- A large anisotropy energy density ($K_{eff}$) value is achieved in the 500°C film along with in-plane UMA (two-fold symmetry).

- In-plane UMA energy remains thermally stable even after annealing the film at 600°C.

- At high $T_A$ of 700°C the Pd/CoFeB/Pd/Ta structure exhibits cubic anisotropy and loses UMA due to excess oxygen diffusion.

**CRediT authorship statement**

**Saravanan Lakshmanan:** Conceptualization, Methodology, Data curation, Investigation, Software, Visualization, Validation, Writing – original draft preparation. **Cristian Romanque:** Formal analysis. **Mario Mary:** Formal analysis. **Manivel Raja Muthuvel:** Data curation. **Nanhe Kumar Gupta:** Data curation. **Carlos Garcia:** Writing – review & editing.

*********************